\documentclass[structabstract]{aa}	
\usepackage{graphicx}

\usepackage{natbib}
\usepackage{color}
\usepackage{url}
\usepackage{txfonts}

\bibpunct{(}{)}{;}{a}{}{,}
\begin{document}

\title{Solar-like oscillations with low amplitude in the CoRoT\thanks{CoRoT (Convection, Rotation and planetary Transits) is a minisatellite developed by the French Space agency CNES in collaboration with the Science Programs of ESA, Austria, Belgium, Brazil, Germany and Spain.}  target HD~181906}
\authorrunning{Garc\'\i a et al.}
\titlerunning{HD~181906}
\author{R. A. Garc\'ia			\inst{1}
	\and C. R\'egulo 			\inst{2,3}
	\and R. Samadi				\inst{4}
	\and J. Ballot 				\inst{5}
	\and C. Barban				\inst{4}
	\and O. Benomar			\inst{6}
	\and W.J. Chaplin  		 	\inst{7}
	\and P. Gaulme			\inst{6}
	\and T. Appourchaux		\inst{6}
	\and S. Mathur				\inst{8}
	\and B. Mosser				\inst{4}
	\and T. Toutain 			\inst{7}
	\and G. A. Verner			\inst{9}
	\and M. Auvergne			\inst{4}
	\and A. Baglin				\inst{4}
	\and F. Baudin				\inst{6}
	\and P. Boumier			\inst{6}
	\and H. Bruntt				\inst{4,10}
        \and C. Catala				\inst{4}
        \and S. Deheuvels			\inst{4}
	\and	Y. Elsworth			\inst{7}
	\and	S.J. Jim\'enez-Reyes 	\inst{2}
	\and E. Michel 				\inst{4}
	\and F. P\'erez Hern\'andez 	\inst{2,3}
	\and I. W. Roxburgh			\inst{9,4}
	\and D. Salabert 			\inst{2}
	}

\institute{Laboratoire AIM, CEA/DSM-CNRS-Universit\'e Paris Diderot; CEA, IRFU, SAp, centre de Saclay, F-91191, Gif-sur-Yvette, France
	\email{Rafael.Garcia@cea.fr}
        \and Instituto de Astrof\'\i sica de Canarias, 38205, La Laguna, Tenerife, Spain
        \and Universidad de La Laguna, 38206 La Laguna, Tenerife, Spain
        \and LESIA, UMR8109, Universit\'e Pierre et Marie Curie, Universit\'e Denis Diderot, Obs. de Paris, 92195 Meudon Cedex, France
        \and Laboratoire d'Astrophysique de Toulouse-Tarbes, Universit\'e de Toulouse, CNRS, 14 av. Edouard Belin, F-31400 Toulouse, France 
        \and Institut d'Astrophysique Spatiale, UMR8617, Universit\'e Paris XI, Batiment 121, 91405 Orsay Cedex, France
        \and School of Physics and Astronomy, University of Birmingham, Edgbaston, Birmingham B15 2TT, UK
        \and Indian Institute of Astrophysics, Bangalore, India 
        \and Astronomy Unit, Queen Mary, University of London Mile End Road, London E1 4NS, UK5 
        \and Sydney Institute for Astronomy, School of Physics, The University of Sydney, NSW 2006, Australia
        }

\date{23 February 2009; 2 June 2009 }

\abstract
{The F8 star HD~181906 (effective temperature $\sim 6300$~K) was observed for 156 days by the CoRoT satellite during the first long run in the centre direction. Analysis of the data reveals a spectrum of solar-like acoustic oscillations. However, the faintness of the target ($m_v$=7.65) means the signal-to-noise (S/N) in the acoustic modes is quite low, and this low S/N leads to complications in the analysis.}
{To extract global variables of the star as well as key parameters of the p modes observed in the power spectrum of the lightcurve.}
{The power spectrum of the lightcurve, a wavelet transform and spot fitting have been used to obtain the average rotation rate of the star and its inclination angle. Then, the autocorrelation of the power spectrum and the power spectrum of the power spectrum were used to properly determine the large separation. Finally, estimations of the mode parameters have been done by maximizing the likelihood of a global fit, where several modes were fit simultaneously.     }
{We have been able to infer the mean surface rotation rate of the star ($\sim$4 $\mu$Hz) with indications of the presence of surface differential rotation, the large separation of the p modes ($\sim$87 $\mu$Hz), and therefore also the ``ridges" corresponding to overtones of the acoustic modes.}
{}


\maketitle

\section{Introduction}
\label{Intro}

CoRoT (Convection, Rotation and planetary Transits) is a minisatellite launched on December 26, 2006 and developed by the French Space agency CNES (Centre National d'Etudes Spatiales) in collaboration with the Science Programmes of ESA (European Space Agency), Austria, Belgium, Brazil, Germany and Spain. The main objectives are to detect exoplanets and to probe the interiors of stars using  asteroseismology \citep{2006cosp...36.3749B}. The high-perfomance photometric data offer an unprecedented opportunity to detect extremely low amplitude oscillations in many stars \citep{2008CoAst.156...73M}, in particular, in solar-like oscillating stars. The main programme of the CoRoT mission consists of 5-month-long runs during which ten designated targets are almost continuously monitored with a 32-s cadence for seismic studies. Several solar-like oscillation stars have already been observed with CoRoT \citep{2008CoAst.157..288G}. One of them is HD~181906, a faint F8 star which has been observed during the first long run looking into the galactic centre direction at the same time as HD~181420 \citep{Barba09}.

In the present paper  we report the first detailed seismic analysis of HD~181906 using CoRoT data. We first present some classical characteristics of the star and infer some expected pulsation properties in Sect.~\ref{Sec:2}; after a description of the CoRoT observations (Sect. 3), we derive the surface rotation of the star (Sect. 4), we then determine the region where the p modes are (Sect. 5) and we explain how to take into account the convective background (Sect. 6). We finish by analysing the p-mode spectrum (Sect. 7) and extracting the p-mode parameters (Sect.8).

\section{Stellar parameters and estimated oscillation properties}\label{Sec:2}

The star HD~181906 (or HIP~95221) is known as an F8 dwarf with a magnitude $m_v=7.65$ and is probably a binary system as suggested by astrometric measurements \citep{2005AJ....129.2420M, 2007A&A...464..377F} or high-resolution spectroscopy \citep{Bruntt2009}. The new reduction of Hipparcos data by \citet{2007A&A...474..653V} provides a parallax $\pi = 14.72 \pm 0.91~\mathrm{mas}$ (i.e. a distance $d=68\pm 4~\mathrm{pc}$), which leads to an absolute magnitude of $M_V=3.49\pm 0.13$. Using BC$_V$${}=-0.044 \pm 0.058$ for the bolometric correction in the V band \citep{2006A&A...450..735M}, we deduce  the stellar luminosity $L= 3.32 \pm 0.45~L_{\sun}$.
Very recently \citet{Bruntt2009} has derived global parameters for this star from high-resolution spectroscopy. He found a value of $T_\mathrm{eff}=6300 \pm 150$~K which is in perfect agreement with the value he derived from 2MASS photometry ($T_\mathrm{eff}=6360 \pm 100$~K). He also deduced a surface gravity $\log g = 4.220 \pm 0.056$ and a metallicity $[\element{Fe}/\element{H}]$ = $-0.11\pm$ 0.14. Using his observations as inputs for stellar modelling, he estimated for this star a mass $M=1.144 \pm 0.119~\mathrm{M}_{\sun}$, a radius $R = 1.392 \pm 0.054~\mathrm{R}_{\sun}$ and an age of $4.2 \pm 1.6$~Gyr. These values have been hereafter used to approximate seismic quantities and are summarized Table~\ref{Tab:quantities}.

\begin{table}[!ht]\centering  
  \caption{Global parameters of HD~181906 used for this work.}
  \label{Tab:quantities}
  \begin{tabular}{lr}
    \hline\hline
    $\pi$                   &  $14.72 \pm 0.91~\mathrm{mas}$ \\
    $L/L_{\sun}$        &  $3.32 \pm 0.45$\\
    $T_\mathrm{eff}$         &  $6300 \pm 150$~K \\
    $\log g$                &  $4.220 \pm 0.056$ \\
    $[\element{Fe}/\element{H}]$ & $-0.11 \pm 0.14$ dex\\
    $v \sin i$              &  $10 \pm 1~\mathrm{km\,s^{-1}}$\\
    \hline
    $M/M_{\sun}$             &  $1.144 \pm 0.119$ \\
    $R/R_{\sun}$             &  $1.392 \pm 0.054$ \\
    \hline
  \end{tabular}
\end{table}

We have compared these values to those derived in previous works using differents techniques.
This star is included in the Geneva-Copenhagen survey of the Solar neighbourhood (\citet{2004A&A...418..989N} revisited by \citet{2007A&A...475..519H}). This survey, mainly based on Str\"omgren photometry, provides an effective temperature $T_{\mathrm{eff}}=6382\pm 91$~K and metallicity $[\element{Fe}/\element{H}] =-0.18\pm 0.10$~dex. The age is also estimated to $2.7^{+0.3}_{-0.4}$~ Gyr and the mass to $1.22^{+0.06}_{-0.07}$~M${}_{\odot}$. All of these values are in very good agreement with those of \citet{Bruntt2009}.
By using V and 2MASS IR photometry, \citet{2006A&A...450..735M} recover a slightly higher effective temperature of $6532\pm 66$~K and provide an angular semi-diameter SD${}=0.102\pm 0.001$\ mas. Thus, using the new Hipparcos parallax, we estimate the stellar radius to $1.50\pm 0.10~R_{\sun}$\footnote{this number is 5\% lower that the one derived by \citet{2006A&A...450..735M}, since the parallax has been revised, but is fully compatible within the error bar.}, which is consistent with \citet{Bruntt2009}.

For the the rotation velocity, \citet{2004A&A...418..989N} have reported a measurement based on earlier CORAVEL observations and have estimated the rotation velocity $v\sin i \approx 16 \pm 1~\mathrm{km\,s^{-1}}$. However, according to  \citet{Bruntt2009}, this value is overestimated due to the blend with a second star. By taking into account the presence of this second star, $v\sin i$ decreases to $10 \pm 1~\mathrm{km\,s^{-1}}$. For consistency reasons, we have considered the latter.


Using these parameters, we are able to obtain approximate values of some global seismic parameters using scaling laws.
The large separation, $\Delta \nu$, can be estimated from the mass and the radius as follows \citep{1995A&A...293...87K}:
\begin{equation}
\Delta \nu = \left(\frac{M}{M_{\sun}} \right)^{1/2} \cdot \left( \frac{R}{R_{\sun}} \right)^{-3/2} \cdot 135 \; \mathrm{\mu Hz}  \; .
\end{equation}
A value of $\Delta\nu\approx 88 \pm 7\; \mathrm{\mu Hz}$ has been obtained.

The frequency of the maximum of the p modes scales as the cut-off frequency \citep{1995A&A...293...87K}. This law has lately been  verified using spectrometric data \citep{2003PASA...20..203B}. Therefore we can express $\nu_{max}$ as follows:
\begin{equation}
 \nu_{max}  = \frac{M/M_{\sun}}{(R/R_{\sun})^2 \cdot  \sqrt{\frac{T_{\mathrm{eff}}}{5777}}}  \cdot 3050 \; \mathrm{\mu Hz} \; .
\end{equation}
We obtain $\nu_{max} \approx 1725 \pm 225\; \mathrm{\mu Hz}$

Finally, the maximum expected amplitude (an estimate of the intrinsic mode amplitude in terms of bolometric intensity fluctuations) can be deduced from \citet{2007A&A...463..297S}:
\begin{equation}
A_{max} = \left( \frac{dL}{L} \right)_{max} = \left(\frac{L/L_{\sun}}{M/M_{\sun}}\right)^{0.7} \cdot \sqrt{\frac{5777}{\mathrm{T_{eff}}}} \cdot A_{\sun \, max} 
\end{equation}
where we have changed the solar value of  $A_{\odot \, max}=$ 2.6 ppm to  2.53 $\pm$ 0.11 ppm  obtained by  calibrating different helioseismic measurements of the VIRGO/SoHO package (POM6 \& SPM) \citep{2009A&A...495..979M}. This gives a value of $A_\mathrm{max}\approx  5.1 \pm 0.6\ \mathrm{ppm}$ for HD~181906. This relation is the combination of two different ones:  the adiabatic relation proposed by \citet{1995A&A...293...87K} to relate mode amplitudes in intensity to mode amplitudes in velocity and the scaling law proposed by  \citet{2007A&A...463..297S} which gives the mode amplitudes in velocity as a function of $(L/M)$.

\section{CoRoT Observations}
\label{Obs}

\begin{figure}[!htb]		
\includegraphics[width=0.47\textwidth]{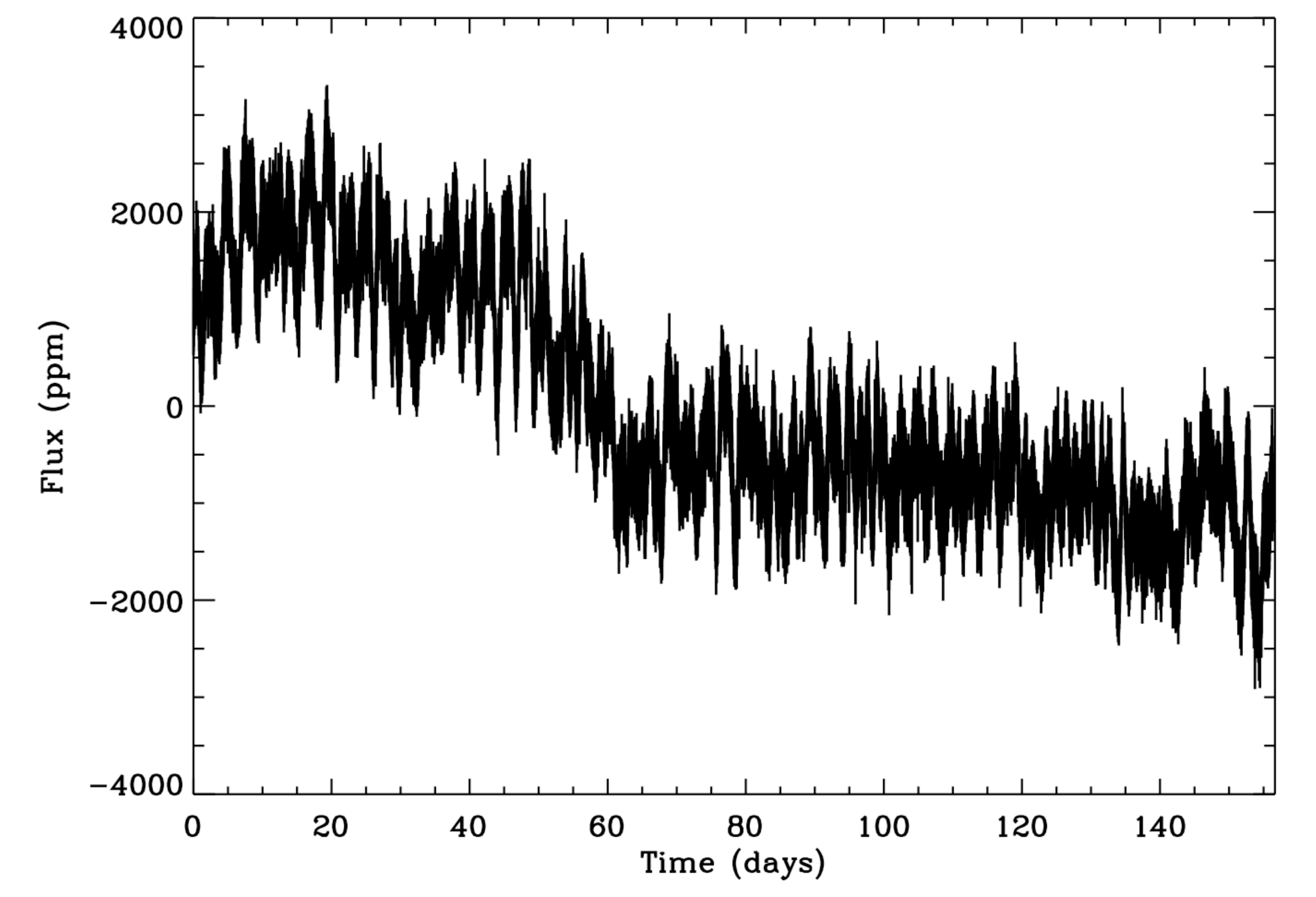}
\caption{N2-Light curve (in ppm) corrected for the aging of the CCD and interpolated onto a regular grid in the heliocentric frame.  }
\label{lightcurve}
\end{figure}

156.6 days of continuous data  -- from 2007 May 11 until 2007 October 15 --  have been collected with an overall duty cycle of 89.3\,\%. Most of the gaps (each of a few minutes duration) are due to data loss during the crossings of the South Atlantic Anomaly \citep[see for a detailed explanation:][]{2009arXiv0901.2206A}.  These gaps have been linearly interpolated in the light curve to avoid putting zeros. We have verified that this interpolation does not introduce any spurious frequencies in the Fourier domain. The original 1-s cadence raw data have been corrected and calibrated into level-2 (or N2) data following the methods described in \citet{2007astro.ph..3354S}. The N2 light curve is sampled on a regular grid in the heliocentric frame with a cadence of 32s. Then we have removed a low-frequency trend due to the aging of the CCD \citep{2009arXiv0901.2206A}, and finally we have removed some outlier points (0.013\%). The resultant lightcurve shows a small modulation during the first 60 days (see Fig.~\ref{lightcurve}) and then a very flat behavior. At this point, it is not possible to disentangle a real modulation of the star from an instrumental effect. A second modulation  -- of around 3 days --  is also visible and could be related to the surface rotation due to the signature of magnetic activity on the surface of the star.

\begin{figure}[!htb]		
\includegraphics[width=0.47\textwidth]{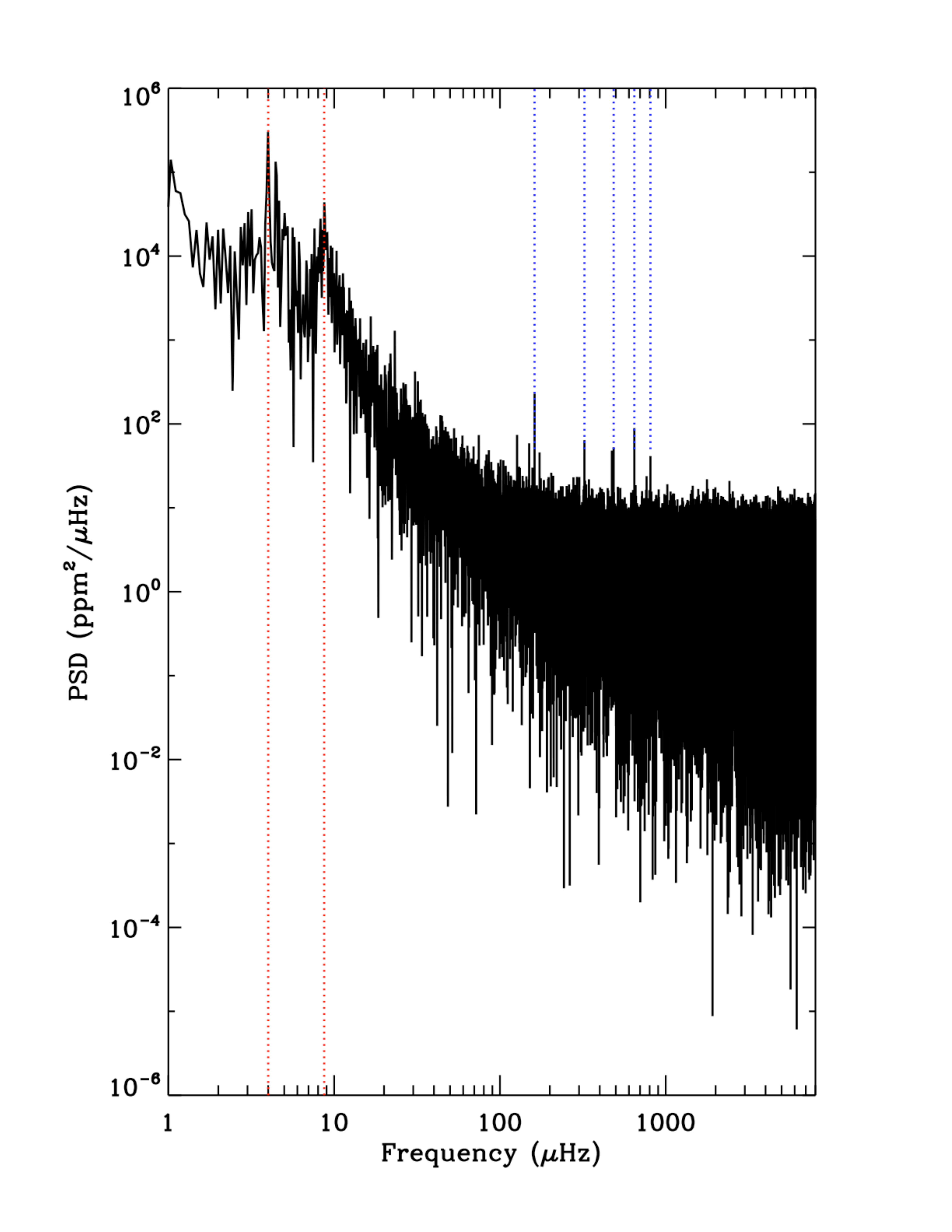}
\caption{Power spectrum density of the full 156-day N2-light curve shown in Fig~\ref{lightcurve}. The two red-dotted vertical lines show the lowest significant peak and the first harmonic that could be the signature of the surface rotation of the star. The five blue-dotted vertical lines indicate the first harmonics of the orbital period of the satellite.}
\label{psdfull}
\end{figure}

To compute the power spectrum density (PSD) we have used a standard Fast Fourier Transform algorithm and normalized it as the so-called one-sided power spectral density \citep{1992nrfa.book.....P}. The resultant PSD is shown in Fig.~\ref{psdfull}. 

In the PSD, several peaks rise above the general trend dominated by the photon noise from the Nyquist frequency down to approximately 100 $\mu$Hz, and by the convective noise from this frequency until $\sim$10 $\mu$Hz \citep{2008Sci...322..558M}. Below this, the spectrum is dominated by two peaks, at around 4 and 8 $\mu$Hz, that could be the signature of the surface rotation of the star  -- already seen in its lightcurve --  as well as its first harmonic (red dotted lines in Fig~\ref{psdfull}). We will discuss in detail this rotation rate in the next section. The signature of the CoRoT orbital periodicity produces a peak at 161.7 $\mu$Hz together with several harmonics (blue dotted lines in Fig.~\ref{psdfull}).  

The combination of a lower-than-expected signal-to-noise ratio of the oscillation amplitudes \citep{2008Sci...322..558M} with the faintness of the target ($m_v$=7.65) means that the p-mode hump is not clearly visible in Fig.~\ref{psdfull}, but there is a small excess in power around $\sim$2000 $\mu$Hz. A more sophisticated treatment is necessary to clearly unveil the acoustic spectrum of this star.

\section{Surface rotation}

As we have already mentioned in previous sections, the light curve of HD~181906 shows a periodic modulation of about 3 days that produces two peaks in the PSD. It is interesting to analyse this periodicity in a more detailed way.

\begin{figure}[!htb]		
\includegraphics[width=0.47\textwidth,height=4.cm,clip]{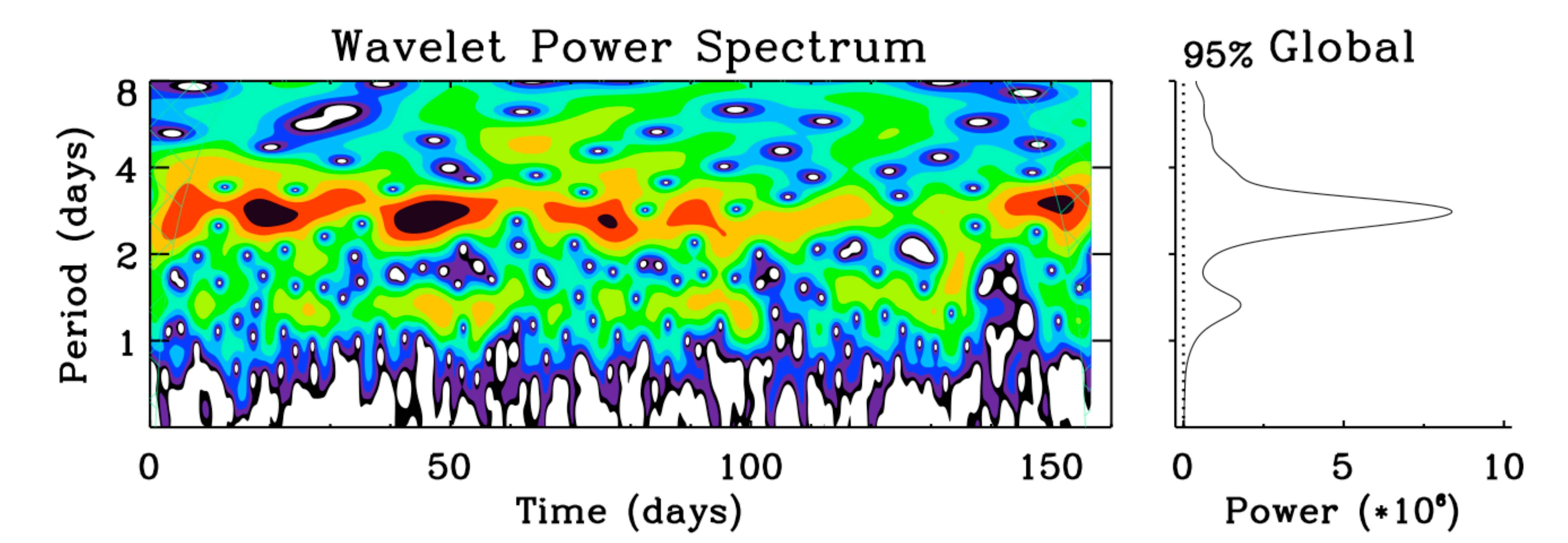}
\caption{Left: Wavelet power spectrum for HD~181906 at low frequency. Right: Global Wavelet Power Spectrum where the dotted-line represents the 95\% confidence level.}
\label{wvrota}
\end{figure}

We have started by calculating a time-period diagram using wavelets \citep{1998BAMS...79...61T}. The advantage of this technique relies on the fact that we use a wave, the Morlet wavelet  -- a sine wave modulated by a Gaussian \citep{Goupillaud84cycle} --  which has a finite duration and a specific frequency. By changing the frequency of this wavelet and sliding it along our time series, we calculate the correlation between the wavelet and the data. That enables us to compute the Wavelet Power Spectrum (see Fig.~\ref{wvrota}). Most of the power is concentrated along a horizontal line centered at $\sim$ 2.8 days. This signal appears to be stronger during the first half of the run than during the second period.
 
With the Global Wavelet Power Spectrum defined as the horizontal average of the time-period diagram (see Fig.~\ref{wvrota} {\it(right)}), we observe that most of the power  -- more than 99\% --  is concentrated in this main periodicity at 2.81 days. A smaller peak is also visible at 1.41 days but containing much less power.  This method (successfully tested with numerical simulations and with solar data from the GOLF instrument \citep{2008arXiv0810.1803M}) allows us to disentangle between the peak corresponding to the main periodicity and that of the harmonic by the simple visual inspection of the PSD where 2 peaks of similar characteristics stand at 4.04 and 8.2 $\mu$Hz.

\begin{figure}[!htb]		
\includegraphics[width=0.5\textwidth,trim=0 2cm 0  2.6cm,clip]{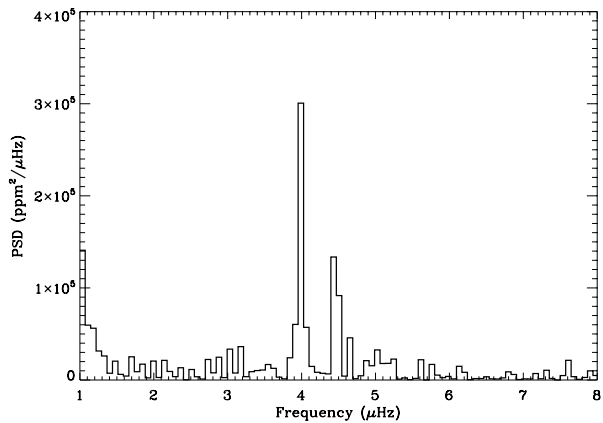}
\caption{Zoom on the spectrum at low frequency. The power spectrum density has been computed using the total length of data.}
\label{figt}
\end{figure}

A closer inspection of the rotation period, $P_{rot}$, in the PSD reveals that it is composed of a double structure, with a strong peak centered at 4 $\mu$Hz  -- 2.9 days (Fig.~\ref{figt}) --  and a smaller one at 4.45 $\mu$Hz (2.6 days). The fact of having these two peaks instead of just one may suggest the presence of spots at different latitudes with a differential surface rotation. The spot modeling done by \citet{2009MosserBaudinLanza} states this point explicitly, indicating two rotation periods associated with a clear gradient of the rotational velocity as a function of spot latitude. Another explanation, less probable, is the presence of a hot star in the background of HD~181906, as suggested by \citet{Bruntt2009} who proposed a detailed analysis of the fundamental parameters of CoRoT asteroseismic targets based on high-resolution spectrometric measurements.

\begin{figure}[!htb]		
\includegraphics[width=0.5\textwidth,trim=0 2cm 0  2.6cm,clip]{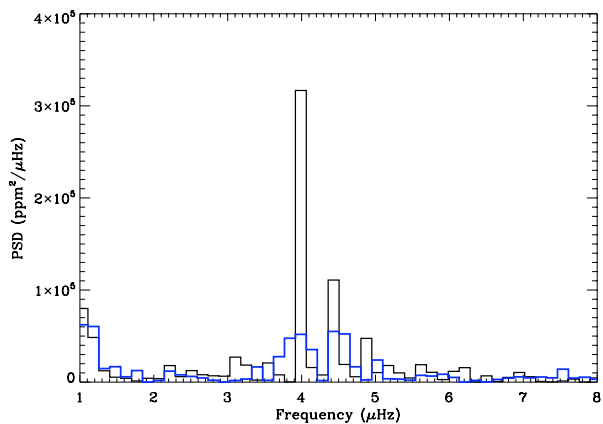}

\caption{Zoom on the spectrum at low frequency. In black the Fourier spectrum of the first 78 days of data. In blue the spectrum of the last 78 days.}
\label{fig78}
\end{figure}

\begin{figure}[!htb]		
\includegraphics[width=0.5\textwidth,trim=0 2cm 0  2.6cm,clip]{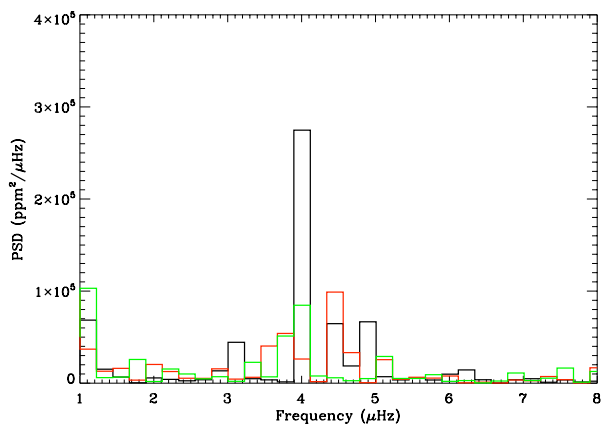}
\caption{Zoom on the spectrum at low frequency. In black the Fourier spectrum of the first 52 days of data. In red the spectrum of the 52 intermediate days of data, and in green the spectrum of the last 52 days.}
\label{fig52}
\end{figure}

If the 156-day time series is divided into two segments of 78 days, or into three independent 52-day subsets, a slightly different low-frequency structure is observed each time, as can be seen in Fig.~\ref{fig78} and Fig.~\ref{fig52}.

In Fig.~\ref{fig78}, the PSD of the first 78 days is plotted in black, and the PSD of the last 78 days is plotted in blue. In Fig.~\ref{fig52}, the PSD of the first 52 days is plotted in black, the next 52 days is plotted in red, and the last 52-day segment is plotted in green. Details of the prominent peaks that appear in each of these spectra are summarized in Table~\ref{tab2}. In this table, we listed the frequencies (in $\mu$Hz) and the respective rotational periods (in days) of the peaks that appear at low frequency in the Fourier spectrum of the full-length series and the smaller subseries that we have considered in the analysis. The peaks are located at rotational periods in the range from 2.3 to 3.1 days. The different behavior of the peaks (in frequency and amplitude) -- behavior that depends on the observed period -- suggests that they could be the signatures of differential rotation on the surface of the star, and not the rotation period of the secondary star in the binary system (which would be expected to produce a stable peak in all the considered periods). Therefore, HD~181906 shows a slightly larger differential rotation than the Sun, but of the same order of magnitude.  A more detailed analysis of this behavior will be the topic of a future work.

\begin{table}[!]
\caption{\label{tab2}Length of spectra and corresponding rotational periods.}
\begin{center}
\begin{tabular}{c c c c} \hline 
 Spectrum & $\mu$Hz (days) & $\mu$Hz (days) & $\mu$Hz (days)    \\  \hline
 Total spectrum & 4.00 (2.9) &  4.45 (2.6)   &   --   \\
 First 78 days & 4.00 (2.9)  & 4.45 (2.6) & 4.90 (2.4)   \\
 Second 78 days & 4.00 (2.9)  & 4.58 (2.5) &  5.04 (2.4)   \\
 First 52 days& 4.00 (2.9)  & 4.45 (2.6)   & 4.90 (2.4)    \\ 
 Second 52 days& 3.77 (3.1) &  4.45 (2.6)  &   --  \\ 
 Third 52 days & 4.00 (2.9)  &--   &  5.12 (2.3)       \\ \hline
 
\end{tabular}
\end{center}
\label{tab}
\end{table}

\subsection{Constraining the inclination angle}
The angle of inclination of HD~181906 can be inferred from the measured value of $(v\sin i)_{obs}$ and the surface rotation that we have just derived:

\begin{equation}
\sin i = \frac{(v \sin i)_{obs}}{2\pi R \nu_{rot} }	\; .
\end{equation}

Using a rotation frequency of $\nu_{rot} = 4.0 \pm 0.15 ~\mathrm{\mu Hz}$ and a rotation velocity of $(v \sin i)_{obs} = 10 \pm 1~\mathrm{km\,s^{-1}}$ we obtain an angle of $i= 24 \pm 3\degr$. We recall this value is obtained by taking into account the blend with a second star \citep{Bruntt2009}. For comparison, by considering the older value of
$16 \pm 1~\mathrm{km\,s^{-1}}$, we would obtain an angle of $37.5 \pm 4.5\degr$. From the spot modeling \citep{2009MosserBaudinLanza} a value of $45 \pm 10\degr$ has been found. This parameter is extremely important because it affects the amplitude ratios of the components of a multiplet of non-radial modes and it is strongly connected to the rotational splitting \citep[see, for example,][and references therein]{2008A&A...486..867B}. 

\section{Finding the p-mode region}
\label{find}
\begin{figure}[!htb]	
\includegraphics[width=0.5\textwidth,trim=0 2cm 0  2.6cm,clip]{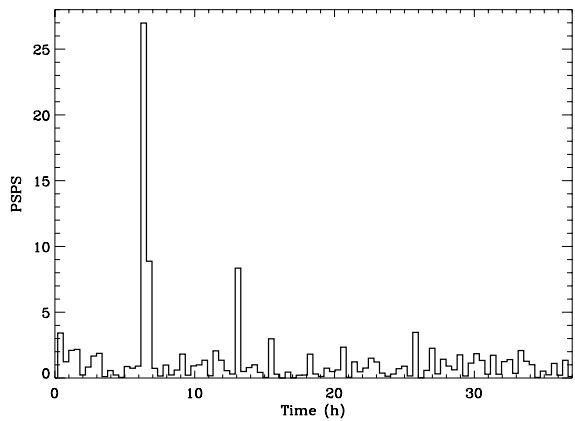}
\caption{Power Spectrum of the Power Spectrum (PSPS) normalized by its standard deviation in the region between 1400 and 2100
$\mu$Hz. The highest peak corresponds to 6.35 hours, i.e. $\Delta \nu$=87.5 $\pm$ 2.6 $\mu$Hz.} 
\label{fig_coin}
\end{figure}

To find the region where the p modes are centered, we computed the
Power Spectrum of the Power Spectrum (PSPS) (or the autocorrelation of
the PSD) in the region determined by the scaling laws (see Fig~\ref{fig_coin} and a more detailed explanation in Sect 7.2). 

This revealed a
strong peak at half the large separation, $\Delta\nu/2$, with the
location of the peak implying $\Delta\nu \sim 87$ $\mu$Hz, which is
close to the upper limit of the value estimated in Section 2, using
existing, non-seismic data on the star. In order to narrow down the
range more carefully, we then performed the following procedure.

We moved a 300-$\mu$Hz-wide window through the frequency range
of interest, and computed the PSPS at each location. The window was
shifted in steps of $33\,\rm \mu Hz$. We measured in each PSPS the
height of the peak at $\Delta\nu/2$ relative to the local background
level. The noise distribution in each PSPS will follow $\chi^2$
two-degrees-of-freedom statistics; given this known distribution, it
is possible to calculate a false-alarm probability for the
$\Delta\nu/2$ peak to appear by chance (as part of the background)
anywhere in the PSPS. We calculated the level corresponding to a 5\%
of probability of appearing by chance \citep{ChaEls2002}.

\begin{figure}[!htb]		
\includegraphics[width=0.5\textwidth,clip]{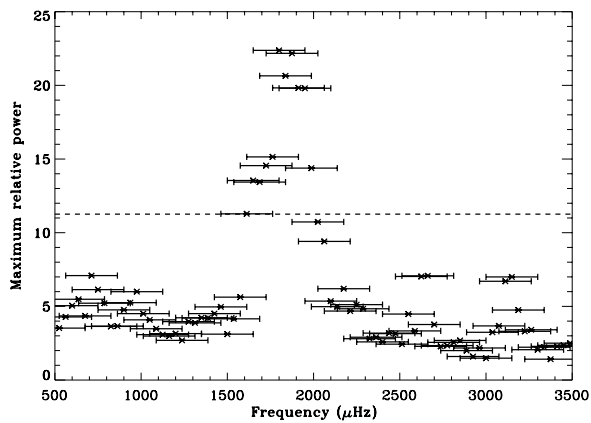}
\caption{Maximum of the PSPS in a region around the expected half-large separation computed in 300 $\mu$Hz-wide windows shifted every 33 $\mu$Hz. The horizontal error bars show the range in frequency in the PS that was analyzed (300 $\mu$Hz). The dashed line represents the 95\% confidence level.}
\label{stat}
\end{figure}

The maximum relative height of the $\Delta\nu/2$ peak (i.e., the
height divided by the measured background in the PSPS) is plotted in
Fig.~\ref{stat}.  We find here that the p-mode power is prominent
between 1400 and 2100 $\mu$Hz.

\section{Background fitting}
\label{Background}
In order to estimate the non-p-mode background, we fit the following three-component model to those parts of the PSD where the observed p-mode power is insignificant ($\mathrm{PSD_{BG}}$):
\begin{equation}
\mathrm{PSD_{\rm BG}(\nu)} = \left(\sum_{i=1}^{2} \frac{A_{i}}{1+(\nu
                    B_{i})^{C_{i}}} \right) + D
\label{eq:bg}
\end{equation}

There are two power-law components in the summation: a component to
represent the significant power at very low frequencies from surface
activity; and a component to represent power from surface granulation.
Both power-law components are modelled in terms of three parameters: a
power spectral density, $A$; a characteristic timescale, $B$; and a
power-law index, $C$. The third component in Equation~\ref{eq:bg},
$D$, models the contribution from shot noise.

Rather than fit the raw power spectrum, we fitted a smoothed spectrum
generated by applying an $N_{\rm s}$-bin-wide boxcar, taking the
independent averages only. With $N_{\rm s} \ga 30$, independent
$N_{\rm s}$-bin averages show normally-distributed scatter about the
(underlying) limit spectrum we seek to estimate from the fit. A
standard least-squares fitting was therefore applied, with weights
fixed by the uncertainties on each independent $N_{\rm s}$-bin
average. These uncertainties were each given by $s/\sqrt{N_{\rm s}}$,
where $s$ is the standard deviation of the $N_{\rm s}$ contributing
power values. 

The alternative approach is to fit the raw spectrum by maximizing a
likelihood function commensurate with the $\chi^2$
two-degrees-of-freedom statistics. It turns out that the smoothed
spectrum may also be fitted by applying the same maximum likelihood
estimator (as was shown by \citep{2004A&A...428.1039A}). These
approaches are, however, more sensitive to the choice of initial
first-guess parameters than is the least-squares fitting approach
applied here, and can as such be prone to poor convergence.

There are a total of seven free parameters defining the background
model in Equation~\ref{eq:bg}. We did not, however, fit them all
simultanously: some parameters were fixed, as we now go on to explain.

We assumed power in the very low-frequency activity component arises
predominantly from the exponential decay of active regions and
plage. This decay implies a limiting power density spectrum that is
Lorentzian, hence we fixed the activity-component index to a value
$C_1=2$ during fitting.

Exercises performed with the asteroFLAG artificial asteroseismology
data \citep{2008AN....329..549C} showed that attempts to fit $A_1$ and
$B_1$ simultaneously with the other parameters could lead to
instability in the procedure and poor convergence. We found that we
could stabilize the fitting by fixing $A_1$ at the value of the power
spectral density of the first element of the averaged spectrum,
leaving $B_1$ as the only parameter of the activity component to be
fitted. This approach does mean that some care is needed in
interpreting the best-fitting value of $B_1$, since the value for
$A_1$ can be affected by the appearence in the power spectrum of
narrow-band features arising from rotational modulation, which are not
modelled in Equation~\ref{eq:bg}. The aforementioned approach to the
fitting does however give a good representation of the power from the
active-region decay that leaks into the frequency region where the
granulation is important; which in turn means we can in principle have
more confidence in the accuracy of the best-fitting granulation
parameters.
\section{The p-mode spectrum}

\begin{figure}[!htb]	
\includegraphics[width=0.47\textwidth]{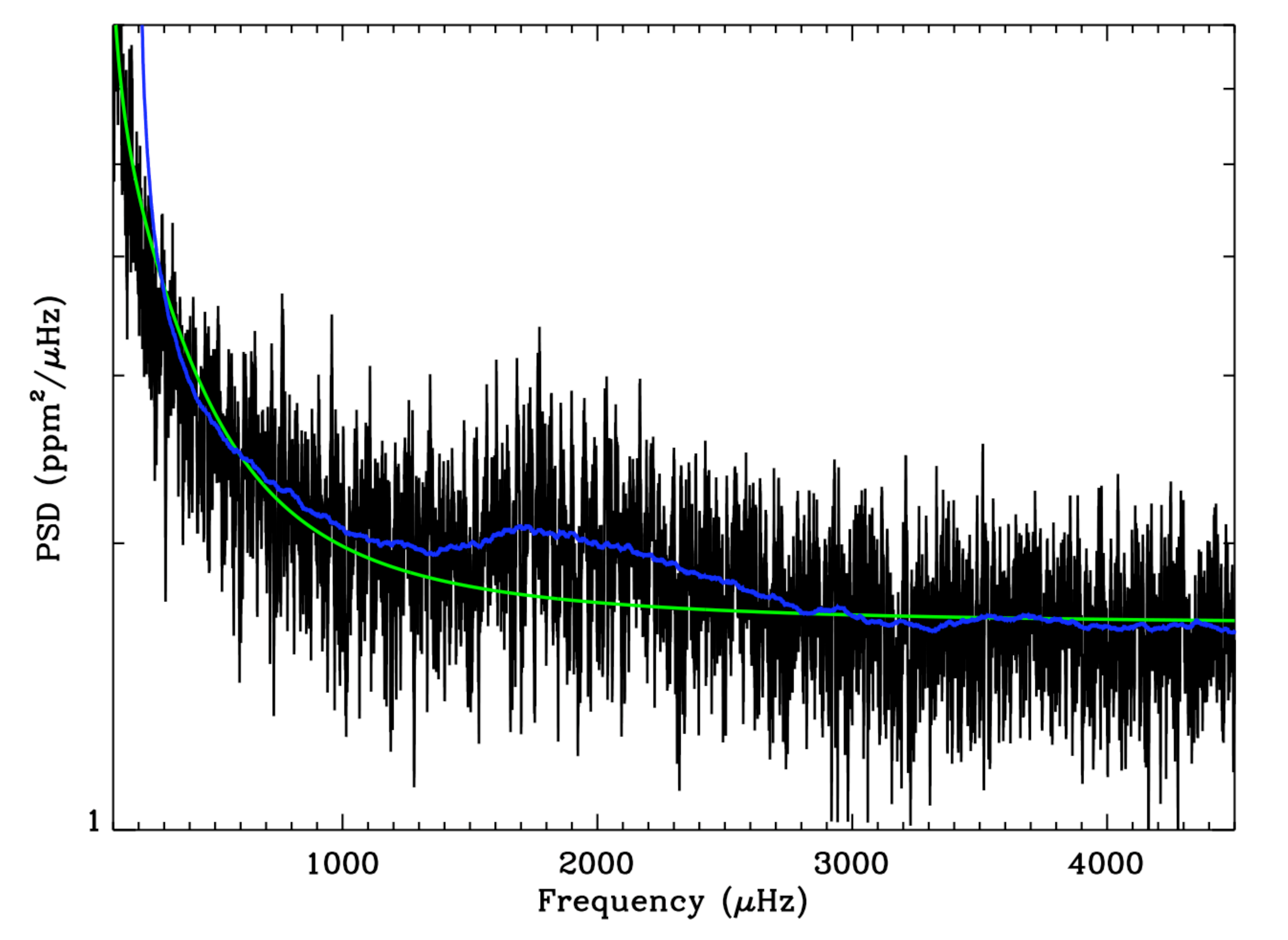}
\includegraphics[width=0.45\textwidth]{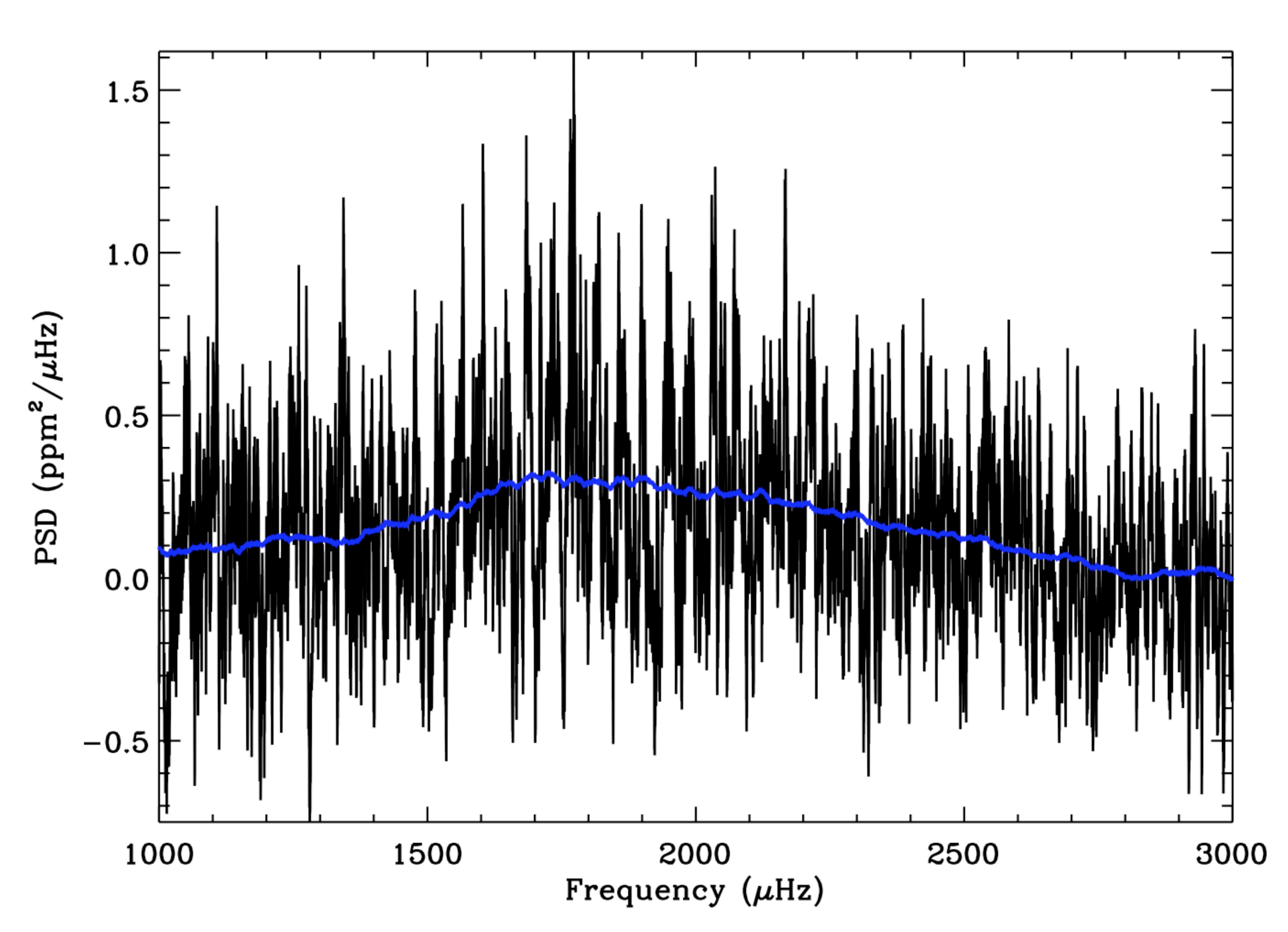}
\caption{Power spectrum density of the full length light curve
smoothed by a boxcar of 70 points (Top). The continuous green curve is the background fitted using the method explained in Section~\ref{Background}.  The bottom panel shows a closer region of the p-mode band between 1000 and 3000 $\mu$Hz after subtracting the background (green curve in the top panel). The blue curve in both panels is the PSD but smoothed using a boxcar of 5000 points.} 
\label{smooth}
\end{figure}

Figure~\ref{smooth} plots the PSD of the full-length lightcurve smoothed
with a boxcar of 70 points ($\sim 5.2$ $\mu$Hz). A clear excess of
power is observed, relative to the best-fitting background model (green
curve), in the region where the p-mode excess was detected (see
Sect.~\ref{find}). To visually enhance the excess we also show in blue
a heavily smoothed spectrum computed by applying a 5000-point boxcar
($\sim 370$ $\mu$Hz) to the raw spectrum.

\begin{figure}[!htb]	
\includegraphics[width=0.47\textwidth]{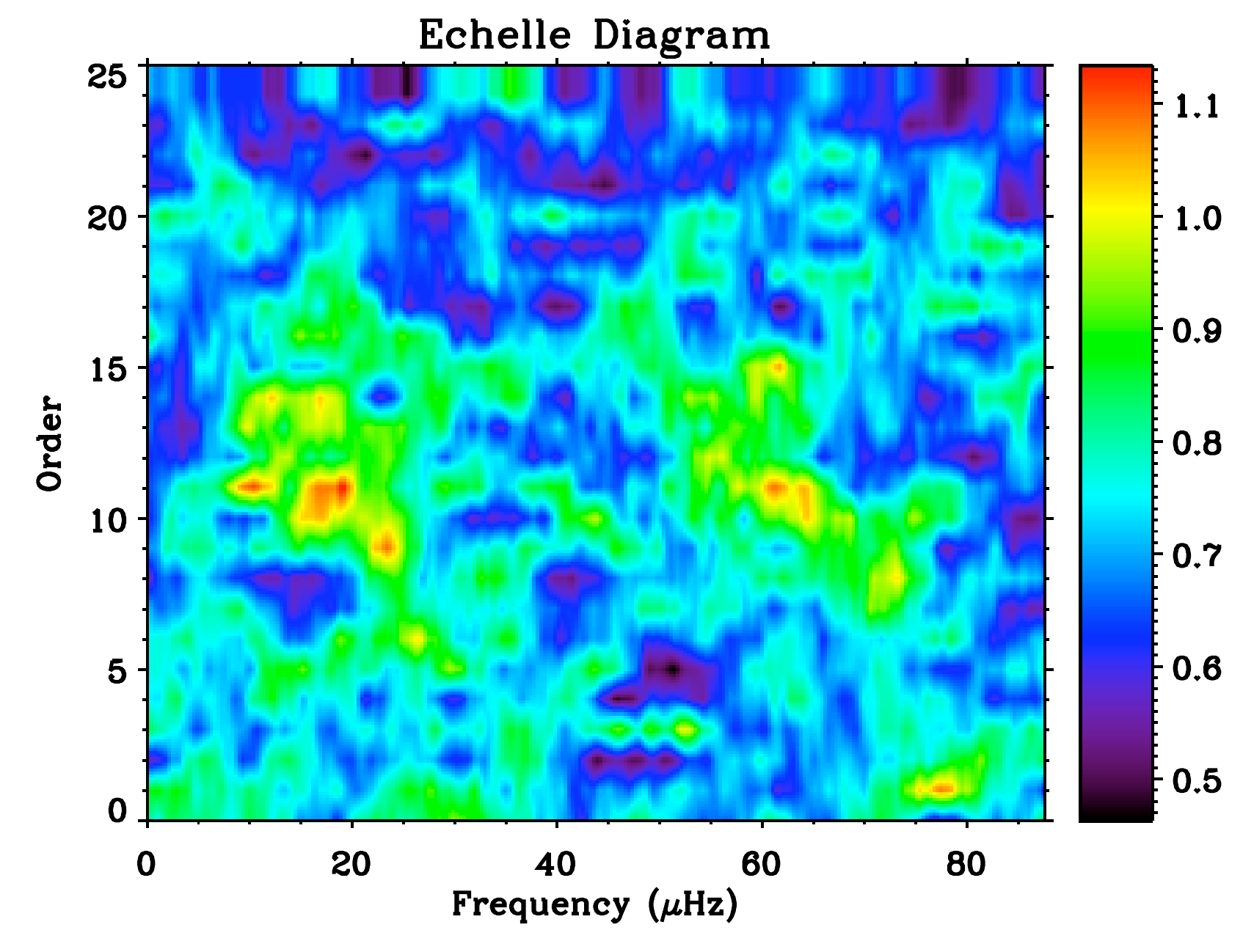}
\caption{Echelle diagram of the full PSD smoothed by a 70 points boxcar starting at a frequency of 792.5 $\mu$Hz and using a folding frequency of 87.5 $\mu$Hz. }
\label{ed}
\end{figure}

Figure~\ref{ed} shows the Echelle diagram \citep{1983SoPh...82...55G}
of the 70-point-boxcar smoothed PSD. The diagram covers the frequency
region from 792.5 to 3000 $\mu$Hz, with each horizontal strip covering
87.5 $\mu$Hz of the spectrum (see Fig.~\ref{fig_coin}). Two ridges appear in the diagram,
corresponding to the odd and even modes. Inspection of the diagram
shows that it is not possible to discriminate visually between the two
ridges, hence the angular-degree tagging is uncertain.

\subsection{Global amplitude}
The procedure for measuring the mode amplitudes begins by averaging
the power spectral density in independent frequency slices of
$q\Delta\nu$, where $q=1$ or 2, and $\Delta\nu$ is the estimated large
frequency spacing of the acoustic mode spectrum. The fitted background
is then subtracted. The resulting residuals in power spectral density
will be greater than zero over ranges occupied by significant mode
power.

Re-calibration of the residual averages then allows for an estimate of
the equivalent $l=0$ mode amplitudes \citep{2008ApJ...682.1370K}. To
re-calibrate, one must: (i) multiply by $q\Delta\nu$, to give the
average \emph{power} across frequency intervals of this length; and
(ii) normalize to power per $l=0$ mode, by dividing by $qV_{\rm tot}$,
where $V_{\rm tot}$ is the combined visibility of the $l=0$, 1, 2 and
3 modes (again, see \citet{2008ApJ...682.1370K}).

In practice we ran a boxcar of width $q\Delta\nu$ through the spectrum
in order to better estimate the maximum power (one could miss the true
maximum if the independent slices cut through pairs of modes). The
measured maximum mode rms amplitude was $A_{\rm max}=2.9 \pm 0.3\,\rm
ppm$.

We also applied a different approach to estimation of the
amplitudes. This approach assumed the envelope of excess power due to
the modes could be modelled by a Gaussian function. We fitted a
Gaussian profile to the re-calibrated residual averages (see above),
and from the best-fitting maximum we estimated $A_{\rm max} = 2.8 \pm
0.1\,\rm ppm$. The location in frequency of maximum also allowed us to
estimate the frequency of maximum power, which was $\nu_{\rm max} =
1912 \pm 47\,\rm \mu Hz$.

To convert these instrumental values into maximum intrinsic bolometric
amplitudes per radial mode, $A_{\rm bol}(l=0$), we took into account
the instrumental response functions for CoRoT, as presented by
\citep{2009A&A...495..979M}. We then found that $A_{\rm bol}(l=0$) =
3.26 $\pm$ 0.42 ppm \citep{2008Sci...322..558M}. This amplitude is
about one third smaller than the expected value deduced from the scaling
laws. Such a deficit is also seen for the other hot, solar-like
oscillators observed by CoRoT \citep{2008Sci...322..558M}. HD~181906 is
not significantly undermetallic, so it is hard to appeal to a
metallicity effect as a possible explanation (e.g., as in
\citet{2008A&A...488..635M}). However, we note that this star -- like
the other solar-like CoRoT targets HD~49933, HD~175726, and HD~181420 --
rotates significantly faster than the Sun (10 times for this target)
and exhibits clear magnetic activity (cf. Sect.~4). It could be an
indication that the magnetic activity plays a role in suppressing the
p-mode amplitudes by changing the surface convection and thus
modifying the efficiency of p-mode excitation processes. Some 3D
compressible radiative magnetohydrodynamics simulations
\citep{2008ApJ...684L..51J} have shown such effects.

There is another possibility, which is that the discrepancy could be
explained by a blending effect. \citet{Bruntt2009} have shown evidence
in the spectrum of HD~181906 for the superimposition of a second
spectrum, which could be a companion or a star in the background
field. If this is the case, the relative amplitudes of modes could
easily be underestimated by a factor of 1.5, making them compatible
with the observations.

\begin{table*}[ht!]
\caption{\label{tab3}Comparison of published global parameters for different F stars.}
\begin{center}
\begin{tabular}{c c c c c c} \hline 
 Stars       & HD181906 & HD49933 & HD181420 & HD175726 & Procyon   \\ 
               &   This paper & Appourchaux et al. 2008 & Barban et al. 2009 & Mosser et al. 2009b  & Arentoft et al. 2008\\ \hline
                    
 Spectral Type & F8     &  F5     &  F2       & F9/G0    &  F5  \\
 $T_\mathrm{eff}$         &  $6300 \pm 150$~K &  6780 $\pm$ 130 K  &  6580 $\pm$ 105 K   & 6000 $\pm$ 100 K  &  6514 $\pm$ 27 K\\
   $[\element{Fe}/\element{H}]$ & $-0.11 \pm 0.14$ dex& $- 0.37$ dex & 0.00 $\pm$ 0.06 dex & $- 0.22\pm$ 0.1 dex &  $-0.05$  dex   \\
 $v \sin i$   &  $10 \pm 1~\mathrm{km\,s^{-1}}$ &    $9.5-10.9 ~\mathrm{km\,s^{-1}}$ & 18 $\pm~1~\mathrm{km\,s^{-1}}$ &  13.5 $\pm~0.5~\mathrm{km\,s^{-1}}$ & 3.16  $\pm~0.5~\mathrm{km\,s^{-1}}$ \\
 $\Delta\nu$ & 87.5 $\pm$ 2.6 $\mu$Hz & 85.9 $\pm$ 0.15 $\mu$Hz & $\sim$ 75  $\mu$Hz  & $\sim$  97$\mu$Hz  & $\sim$ 55  $\mu$Hz  \\ 
 $\nu_{max}$ & 1900  $\mu$Hz & 1760 $\mu$Hz & 1500 $\mu$Hz & 2000  $\mu$Hz & 900  $\mu$Hz \\ 
 A$_{max}$ & 3.26 $\pm$ 0.42 ppm& 4.02 $\pm$ 0.57 ppm &  3.82 $\pm$ 0.40 ppm & $\sim$ 1.7 ppm & $\sim$ 8.5 ppm  \\ \hline
 
\end{tabular}
\end{center}
\label{tab}
\end{table*}
\subsection{The mean large separation}
\label{Large_Sep}

Different methods have been used for extracting the large separation
of the p-mode spectrum. A good signature of this large separation can
be derived from the PSPS, or from the autocorrelation of the power
spectrum, computed in the region where the excess of power has been
detected (see Fig.~\ref{smooth}).

Taking the frequency interval where the p-mode excess has been found
(see Sect.~\ref{find}), i.e. from 1400 to 2100 $\mu$Hz, we find that
the PSPS is dominated by a peak at 6.35 hours (see
Fig.~\ref{fig_coin}), which corresponds to half the large separation
of 43.75 $\mu$Hz. Indeed, the main periodicity we found is not the
large separation itself but the distance between peaks of even and odd
degree. Therefore, the large separation is $\Delta \nu$ = 87.5 $\pm$
2.6 $\mu$Hz (see Fig.~\ref{fig_coin}). Depending on the range in frequency used to look for the
large separation, different groups have found slightly different
values. For example, if we reduce the search range to 1400 to 2000
$\mu$Hz, the large separation is reduced to a value of 85.7 $\pm$ 2.6
$\mu$Hz.

Another powerful way to estimate the value of the large separation is
to build the so-called collapsogram \citep{2008AdSpR..41..897K}. The
first step consists in calculating all the Echelle diagrams for a
frequency range in which we expect the large separation (e.g. 60-100
$\mu$Hz). The second step then consists in collapsing the vertical
dimension of each Echelle diagram. In the last step the collapsed
diagrams are stacked vertically, to yield the final collapsogram.

The collapsogram presents two advantages. First, it allows us to
explore on a single diagram a wide range of possible values of the
large separation. Secondly, it increases the visibility of the
separation compared to that seen in the Echelle diagram (although it
does rely on the separation being almost uniform over the frequency
range used to construct the collapsogram).

\begin{figure}[!htb]	
\includegraphics[width=0.5\textwidth]{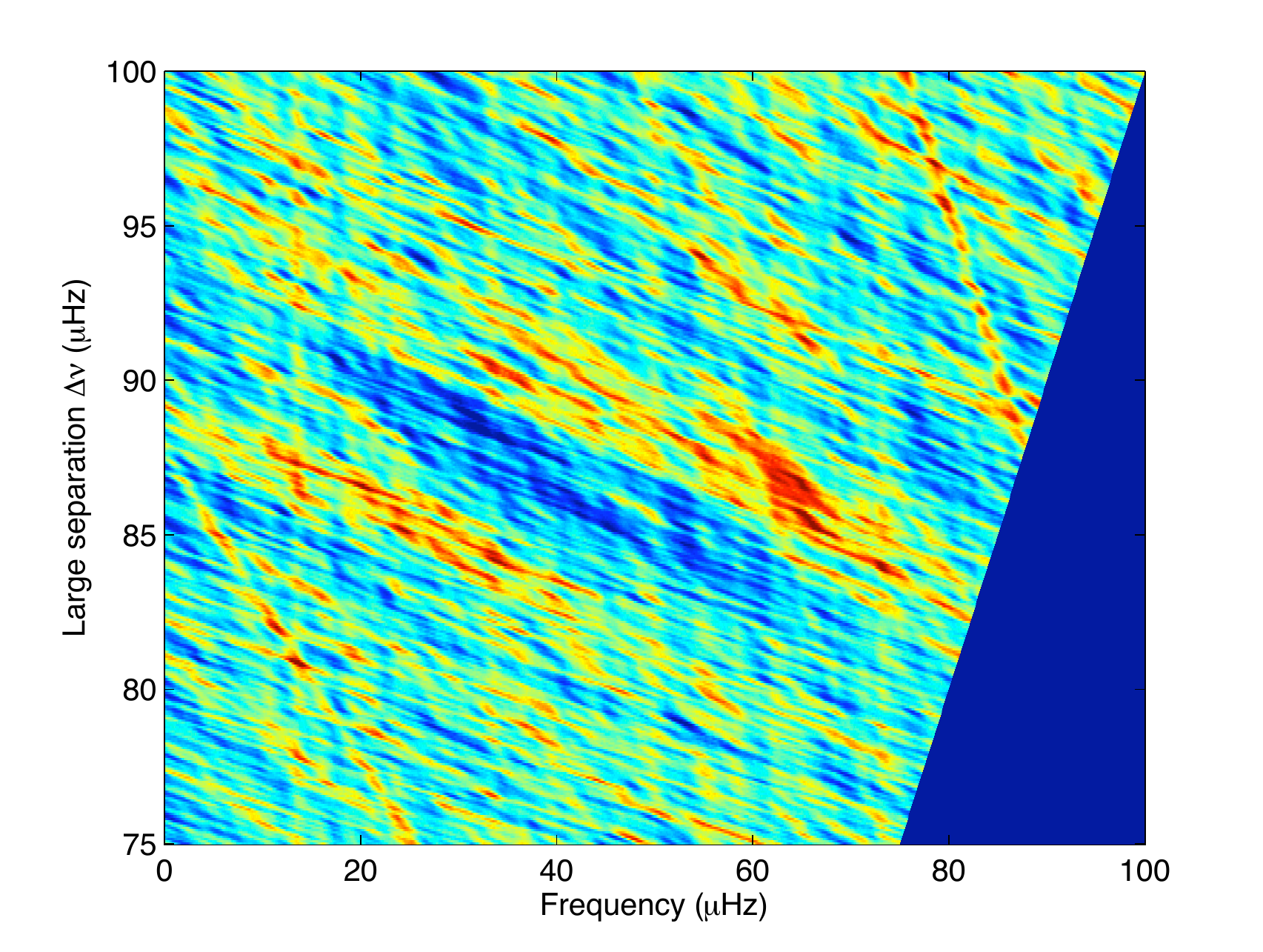}
\caption{Collapsogram of the smoothed power spectrum (11 bins)
obtained by collapsing the power spectrum in the frequency range
between 1280 and 2300 $\mu$Hz. The vertical axis corresponds to the
large separation that was explored, i.e., from 75 to 100 $\mu$Hz. The
horizontal axis indicates the frequency, from 0 to 100 $\mu$Hz, which
is the maximum large separation that was looked for. Note that there
are two prominent excesses of power in the collapsogram, corresponding
to a large separation of about 85 to 90 $\mu$Hz, at $\simeq$27 and
$\simeq$67 $\mu$Hz.}
\label{collapso}
\end{figure}

In Fig.~\ref{collapso}, we present the collapsogram of the smoothed
power spectrum obtained with a weighted moving average over 11
bins. The presence of the two mode ridges is seen, with the large
separation indicated as being between 85 to 90 $\mu$Hz. Note that we
have considered only the power spectrum in the frequency range between
1300 and 2300 $\mu$Hz to build the collapsogram.

\subsection{Comparison with other observed F stars}
\label{Analysis}

A comparison of published global parameters (and their associated error bars when available) from different F stars is presented in Table ~\ref{tab3} where $\Delta\nu$ is the large spacing, $\nu_{max}$  is the frequency in the spectrum with the maximum power and A$_{max}$  is the bolometric maximum amplitude per radial mode. In the case of Procyon, the bolometric amplitude is obtained from the velocity measurements following \citet{2008ApJ...687.1180A}. The stars HD~181906, HD~49933, HD~181420 and HD~175726 are observed by CoRoT. However, the length of observations is not the same for all of them, being 156 days for HD~181906 (the star analyzed in this paper) and HD~181420 \citep{Barba09}, 60 days for HD~49933 \citep{2008A&A...488..705A} and only 27 days for  HD~175726 \citep{2009HD175726}. The data from Procyon correspond to velocity measurements obtained during 26 days for a multisite campaign involving 11 telescopes \citep{2008ApJ...687.1180A}.

The data  in  Table ~\ref{tab} are obtained mainly from the above mentioned papers, except  the bolometric maximum amplitude per radial mode (A$_{max}$) for the 3 first mentioned CoRoT stars that were obtained from \citet{2008Sci...322..558M}. For Procyon, the T$_{eff}$, $v \sin i$, and $[$Fe/H$]$ were obtained from \citet{2002ApJ...567..544A}. The value of the maximum amplitude in the spectrum is a translation from cm/s to ppm done in  \citet{2008ApJ...687.1180A} by the authors, a value that agrees well with the one obtained by \citet{2005ApJ...633..440B} using white-light photometric data from the WIRE satellite.

\section{Extracting individual p-mode characteristics}
\label{sec:globalfitting}
Although fitting low-degree p-mode profiles in helioseismology and asteroseismology might appear very similar,
the unknown stellar inclination angle makes the fitting of asteroseismic data much more difficult (see for example \citet{2008A&A...488..705A} for the case of the star HD~49933 or \citet{2003ApJ...589.1009G} for Monte-Carlo simulations with artificial p-mode profiles).
It is not only the lower signal-to-noise ratio of the p-mode asteroseismic signal which makes the fitting difficult  
but also the high correlation between the inclination and the rotational splitting \citep{BalGar2006,2008A&A...486..867B}.
Because of that, the determination of these two parameters can be rather poor and will consequently
affect the determination of the other parameters (frequencies,
widths, heights...). Therefore, instead of fitting individually each
multiplet or pair of modes -- as it is commonly done in helioseismology -- we choose to perform a global fitting of all the multiplets above a given amplitude threshold around the maximum of the p-mode hump, assuming that the rotational splitting is independent of the frequency \citep[see][for all the details]{2008A&A...488..705A}. This type of global method was pioneered by \citet{1999ESASP.448..135R} using solar data. By doing so, the splitting and the inclination angle are better constrained, even though HD~181906 is then modeled as a rigidly rotating star. Each multiplet is described by five parameters: the central frequencies of the modes $l=0,1,2$, one line width (the same for all modes within a large separation), and one mode height. We assumed the same visibility ratio between angular degrees as the ones used in full-disk helioseismology and the visibilities between $m$-components given by \citet{2003ApJ...589.1009G}.

Eight teams performed a global fitting of the HD~181906 data, seven of them using a maximum likelihood minimization and one using a least square fitting over an averaged spectrum. All the teams did the fits for both scenarios (A and B) depending on the identification of the ridges as the odd or even modes.

\begin{figure}[!htb]
\includegraphics[width=0.5\textwidth]{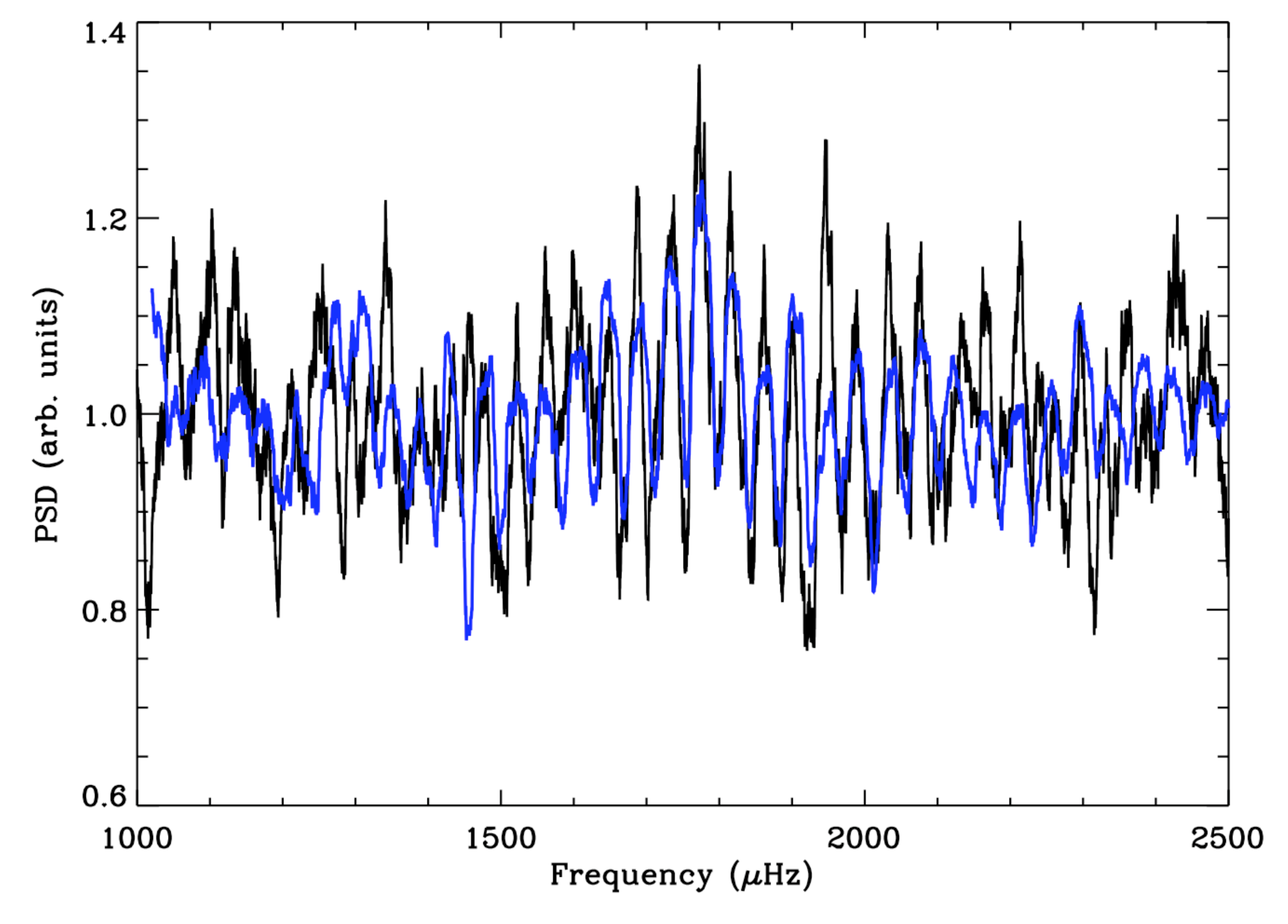}
\caption{Smoothed power spectrum of HD~181906 (black curve) and smoothed and shifted by 17~$\mu$Hz power spectrum of HD~49933 (blue curve).}
\label{fig:49933}
\end{figure}

Two teams decided to split the observations in four subseries and
computed the Joint Power Statistics (JPS, see Appendix A for a
detailed explanation).  The JPS is an alternative method that
contains the same underlying information as the average spectrum but with a different treatment of the noise. The rest of the teams worked on the full resolution power spectrum. The number of overtones fitted was 5, 7, 9 and 16 overtones. Different strategies have also been followed by each team to obtain the initial parameters for the fit. In particular, one team used the results from HD~49933 \citep{2008A&A...488..705A} as a guideline since their PSDs look very similar and HD~49933 has a much better signal-to-noise ratio. Figure \ref{fig:49933} shows the superposition of the PSD of HD~181906 and the one of HD~49933 -- properly scaled in amplitude -- and shifted by 17 $\mu$Hz. Thus, the initial guesses were obtained directly from the fitted values of HD~49933. Another team repeated the fits 200 times for each scenario, adding scatter to the initial parameters in order to test the robustness of the fits and identify any correlation between the initial seed values and the fitted ones. As expected, the inclination angle was identified to be strongly correlated with the initial seed angle.

The mode parameters were then extracted using different approaches
based on the recipe developed by the Data
Analysis Team (DAT) and explained in \citet{2008A&A...488..705A}.
However, in order to have a more reliable characterization of the
p-mode signal, given the low signal-to-noise ratio of the
observations, some more a-priori information needed to be introduced.
In order to stabilize the fits and avoid systematic outliers, some of
the fitting teams fixed some parameters to be equal over the range of
considered overtones. For instance, without some a-priori conditions,
the minimization procedures failed to return reliable estimates at high frequency, fitting mostly spikes and over-estimated heights. Tables~\ref{table:freq_A} and \ref{table:freq_B} show the fitted mode frequencies obtained from the less constrained fit for both scenarios A and B respectively, the analysis being performed on the power spectrum of the whole HD~181906 time series. The only condition applied to those fits is that the mode widths and mode heights were set to be uniform (the same for all the modes) over the fitted frequency range. In the central part of the p-mode hump, those frequency estimates are consistent within $3\sigma$ with the estimates returned using more constrained fits and different dataset lengths.  In the low- and high-frequency ranges, the discrepancies are larger but still within $5\sigma$. Unfortunately, neither of the scenarios seems to be favoured: for instance, the likelihood ratio test does not allow us to disentangle between the two possibilities, the likelihood ratio being not greater than 5 (for the fits presented here). 

\begin{table}[!htb]
\begin{center}
\caption{HD~181906 frequencies (in $\mu$Hz) for scenario A.}
\begin{tabular}{crrr}
\hline\hline
&  & \multicolumn{1}{c}{Scenario A}  & \\
\hline
\multicolumn{1}{c}{\# } & \multicolumn{1}{c}{$l=0$} & \multicolumn{1}{c}{$l=1$} & \multicolumn{1}{c}{$l=2$} \\ 
\hline  
1  &  1524.80~$\pm$~0.86  &  1565.61~$\pm$~0.38  &  1521.98~$\pm$~0.42 \\
2  &  1604.89~$\pm$~0.25  &  1645.98~$\pm$~0.41  &  1596.46~$\pm$~0.51 \\ 
3  &  1684.62~$\pm$~0.40  &  1734.98~$\pm$~0.53  &  1687.32~$\pm$~0.51 \\
4  &  1773.14~$\pm$~0.27  &  1814.29~$\pm$~0.31  &  1772.43~$\pm$~0.39 \\
5  &  1862.25~$\pm$~0.53  &  1898.77~$\pm$~0.27  &  1861.74~$\pm$~0.39 \\
6  &  1947.42~$\pm$~0.30  &  1988.32~$\pm$~0.37  &  1946.26~$\pm$~0.28 \\  
7  &  2035.88~$\pm$~0.62  &  2076.45~$\pm$~0.51  &  2036.30~$\pm$~0.95 \\
\hline
\end{tabular}
\label{table:freq_A}
\end{center}
\end{table}

\begin{table}[!htb]
\begin{center}
\caption{HD~181906 frequencies (in $\mu$Hz) for scenario B.}
\begin{tabular}{crrr}
\hline\hline
&  & \multicolumn{1}{c}{Scenario B} &  \\
\hline
\multicolumn{1}{c}{\# } & \multicolumn{1}{c}{$l=0$} & \multicolumn{1}{c}{$l=1$} & \multicolumn{1}{c}{$l=2$} \\ 
\hline  
1  &  1565.96~$\pm$~0.38  &  1521.94~$\pm$~0.41  &  1560.55~$\pm$~0.65 \\
2  &  1646.63~$\pm$~0.35  &  1604.76~$\pm$~0.29  &  1644.97~$\pm$~0.41 \\
3  &  1734.59~$\pm$~0.26  &  1684.74~$\pm$~0.62  &  1735.54~$\pm$~0.53 \\
4  &  1814.83~$\pm$~0.41  &  1772.68~$\pm$~0.27  &  1814.16~$\pm$~0.41 \\
5  &  1898.79~$\pm$~0.43  &  1861.91~$\pm$~0.43  &  1898.46~$\pm$~0.40 \\
6  &  1988.50~$\pm$~0.33  &  1947.04~$\pm$~0.25  &  1988.72~$\pm$~0.44 \\
7  &  2075.91~$\pm$~0.32  &  2036.48~$\pm$~0.69  &  2077.20~$\pm$~0.38 \\
\hline
\end{tabular}
\label{table:freq_B}
\end{center}
\end{table}

Nevertheless, we can still obtain some global characteristics of
HD~181906. On one hand, the fitted inclination angle, for both scenarios (around
$47^\mathrm{o} \pm 3.5^\mathrm{o}$) , is very close to the one
derived from the analysis of the low-frequency peak of the PSD (see
Sect. 4) when a $(v\sin i)_{obs} \approx 16 \pm 1~\mathrm{km\,s^{-1}}$ is
considered but quite far from the $\sim 24^\mathrm{o}$ obtained considering  $(v\sin i)_{obs} \approx 10~\mathrm{km\,s^{-1}}$. On the other hand, the extracted rotational splitting (5.8 and 6.1
$\pm$ 0.14 $\mu$Hz respectively for A and B scenarios) is
overestimated compared to the surface rotation rate (see Sect. 4).
This could be due to an increase of the rotation rate in the stellar
interior or a biased estimation due to the a-priori conditions
applied to the fit (a common height and width for all the fitted
modes).These results show, once again, the strong correlation between these two parameters and the necessity to have an external good determination of the inclination angle or the rotation of the star to be able to disentangle between the possible solutions.

Another parameter -- the global line width -- was found to be rather small (1.25 -0.3/+0.4 and 1.28 -04/+0.6 $\mu$Hz for scenarios A and B respectively), 
which is probably a consequence of the low SNR that could bias this
estimate when we apply our peak-bagging codes: the low SNR means the
modes will appear ``spikier" than they really are (we lose the
Lorentzian tails in the background noise), meaning we will be biased
to fitting small line widths.  We expect that the  autocorrelation of
the power spectrum will have better SNR than  that seen in individual
modes and hence we look there for the features associated with mode line
width. The features in it are quite wide, implying that the line
widths are large and probably in line with what we saw for HD~49933 (see Fig.~\ref{fig:49933}). Indeed, the widths of the modes in both stars, around the maximum of the p-mode spectrum, seem to be similar which favours shorter mode lifetimes than in the Sun. 

Finally, we would like to emphasize that the combination of a small
SNR, a possible small inclination angle of the star (close to
$25\degr$) and large mode line widths produce similar Lorentzian
profiles for the modes l=0, 2 and 1 \citep{2004ESASP.559..309B}. This
effect could contribute to explaining why we obtain very similar fitting results in both scenarios. More work will be necessary to obtain more reliable individual p-mode parameters.

\section{Conclusion and perspectives}
\label{Conclusion}

In this paper we have shown the first seismic analysis of HD~181906
(HIP 95221), a faint CoRoT target ($m_v$=7.65) that has been observed continuously during 156.6 days in 2007. The surface rotation of the star has been inferred by analyzing the very low-frequency part of the power spectrum. A rotation rate of 2.9 days (4 $\mu$Hz) has been established. The presence of a second peak close to the previous one in the power spectrum and produced during a different period of time, has been interpreted as the signature of the presence of magnetic structures on the stellar surface at different latitudes and, therefore, it has been deduced that this star could have a slightly higher differential rotation than the Sun. Coupling this rotation rate with the previous result of the $(v\sin i)_{obs} \approx 16 \pm 1~\mathrm{km\,s^{-1}}$ has allowed us to infer an inclination angle of $\mathrm{37.5^o}$ $\pm$~$\mathrm{4.5^o}$ but this value could be reduced to $\mathrm{24^o}$ $\pm$~$\mathrm{3^o}$ if we consider that HD~181906 has a companion which implies a reduction in the $(v\sin i)_{obs}$ to  10 $\pm 1~\mathrm{km\,s^{-1}}$.

A comb-like structure has been unveiled between 1400 and 2100 $\mu$Hz
which corresponds to the acoustic-mode spectrum with a maximum power
at 1912 $\pm$ 47 $\mu$Hz and a large separation of 87.5 $\pm$ 2.6
$\mu$Hz measured inside this frequency interval. However, the low
signal-to-noise ratio of the modes prevents us from unambiguously
identifying them. To go further,  more a-priori information is needed
to constrain the fits. 

\begin{acknowledgements}
The CoRoT space mission, launched on December 27th 2006, has been developed and is operated 
by CNES, with the contribution of Austria, Belgium, Brazil , ESA (RSSD and Science Programme), 
Germany and  Spain. J.B. acknowledges support through the ANR project Siroco. I.W.R. and G.A.V. wish to thank the UK Science and Technology Facilities Council for support under grant PP/E001793/1. W.J.C. and Y.E. also wish to thank the UK Science and Technology Facilities Council for support under grant ST/F00204/1. H.B. was supported by the Australian and Danish Research Councils. D.S. acknowledges the support of the Spanish National Research Plan under the grant PNAyA2007-62650. Wavelet software was provided by C. Torrence and G. Compo, and is available at URL: \url{http://atoc.colorado.edu/research/wavelets/}

\end{acknowledgements}

\bibliographystyle{aa}
\bibliography{/Users/rgarcia/Desktop/BIBLIO} 

\appendix 
\section{Joint Power Statistic}

In cases where the signal-to-noise ratio of a power spectrum is low, it can be useful to increase the signal and decrease the noise at a cost of decreasing the resolution in frequency.  This has often been accomplished by calculating either the arithmetic or geometric mean of a set of power spectra obtained from independent contiguous subsets of the complete time series.  The expected statistics of these averaged power spectra can be calculated, however they do not have the same negative exponential distribution as a single power spectrum.  The standard maximum likelihood estimation (MLE) fitting techniques are designed for power spectra with negative exponential statistics (\textit{i.e.} $\chi^{2}$ with 2 d.o.f.), it is therefore desirable to calculate an averaged power spectrum with the same statistics as the original spectrum.

The joint power statistic (JPS) \citep{2005SoPh..227..137S} esembles a correlation function in the sense that it increases the contribution of signals present in a number of independent power spectra while decreasing the uncorrelated noise.  It also has the important property that the JPS is distributed with negative exponential statistics and can, therefore, be immediately fitted using established MLE techniques. The JPS can be calculated using the power spectra of any number of independent subseries, with the resolution in frequency decreasing by a proportional amount compared with the power spectrum of the complete time series.  In the case of HD~181906, the best compromise between increasing the amplitude-to-background ratio and maintaining sufficient resolution in frequency was found when the complete detrended time series was divided into four contiguous subseries.  The appropriate fourth order JPS can be approximated by
\begin{equation}
  J_{4A} = \frac{3.881X^2}{1.269+X},
  \label{eq:jps} 
\end{equation}
where $X$ is the geometric mean of the power spectra ($S_i$) calculated from four independent subseries
\begin{equation}
   X = (S_1S_2S_3S_4)^{1/4}.
\end{equation}

In the JPS calculated from four subseries, after smoothing with a Gaussian filter of width 1$\sigma$=2\,$\mu$Hz, the average amplitude-to-background ratio for mode peaks between 1700 and 2000\,$\mu$Hz is 2.8.  This compares with the equivalent full series smoothed FFT amplitude-to-background ratio of 1.7 over the same range.  While the JPS can be fitted assuming negative exponential statistics, the mode amplitudes obtained will be higher than those obtained from a single power spectrum due to the increase in power of correlated signals in the JPS.  Similarly, the stellar background parameters obtained from a JPS fit will be lower than those obtained from a power spectrum fit.  The mode frequencies and linewidths obtained from the JPS are the only parameters that are directly comparable with a fit to the power spectrum.

\end{document}